\documentclass[11pt]{article}
\usepackage{hyperref}
\pdfoutput=1
\begin{document}
\title{Vortices around Dragonfly Wings \\ Fluid Dynamics Videos}
\author{Jihoon Kweon and Haecheon Choi \\
\\\vspace{6pt} School of Mechanical and Aerospace Engineering, \\ Seoul National University, Seoul 151-744, Korea}
\maketitle
\begin{abstract}
Dragonfly beats its wings independently, resulting in its superior maneuverability. Depending on the magnitude of phase difference between the fore- and hind-wings of dragonfly, the vortical structures and their interaction with wings become significantly changed, and so does the aerodynamic performance. In this study, we consider hovering flights of modelled dragonfly with three different phase differences ($\phi= -90^\circ, 90^\circ, 180^\circ$). The three-dimensional wing shape is based on that of \emph{Aeschna juncea} (Norberg, 1972), and the Reynolds number is 1,000 based on the maximum translational velocity and mean chord length. The numerical method is based on an immersed boundary method (Kim \emph{et al}., 2001). In counter-stroke ($\phi=180^\circ$), the wing-tip vortices from both wings are connected in the wake, generating an entangled wing-tip vortex (e-WTV). A strong downward motion induced by this vortex decreases the lift force in the following downstroke (Kweon and Choi, 2008). When the fore-wing leads the hind-wing ($\phi=-90^\circ$), the hind-wing is submerged in the vortices generated by the fore-wing and suffers from their induced downwash flow throughout the downstroke, resulting in a significant reduction of lift force. On the other hand, when the hind-wing leads the fore-wing ($\phi=90^\circ$), the e-WTV is found only near the start of hind-wing upstroke. In the following downstroke of hind-wing, most of the e-WTV disappears and the hind-wing is little affected by this vortex, which produces relatively large lift force.

\end{abstract}
\section{Introduction}
Two videos are
\href{http://ecommons.library.cornell.edu/bitstream/1813/14084/3/Vortices_around_Dragonfly_Wings_MPEG1.mpg}{MPEG1} for display at the meeting and 
\href{http://ecommons.library.cornell.edu/bitstream/1813/14084/2/Vortices_around_Dragonfly_Wings_MPEG2.mpg}{MPEG2} for display at DFD website.

\end{document}